# Differentiation of Complex Vapor Mixtures Using Versatile DNA-Carbon Nanotube Chemical Sensor Arrays

Nicholas J. Kybert,[1] Mitchell B. Lerner,[1] Jeremy S. Yodh,[1] George Preti,[2,3] and A. T. Charlie Johnson[1,*]

[1] Department of Physics and Astronomy and Nano/Bio Interface Center, University of Pennsylvania, Philadelphia, Pennsylvania  19104, United States

[2] Monell Chemical Senses Center, Philadelphia, Pennsylvania  19104, United States

[3] Department of Dermatology, School of Medicine, University of Pennsylvania, United States

* cjohnson@physics.upenn.edu

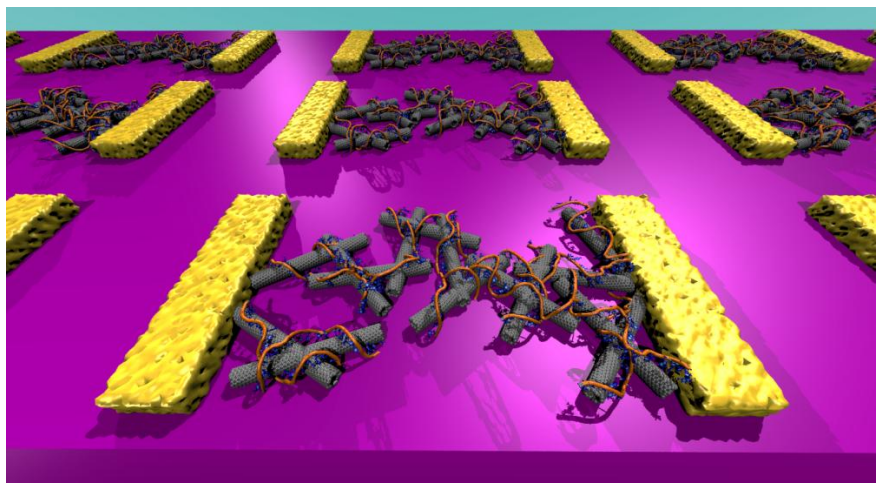

**Abstract**

Vapor sensors based on functionalized carbon nanotubes (NTs) have shown great promise, with high sensitivity conferred by the reduced dimensionality and exceptional electronic properties of the NT. Critical challenges in the development of NT-based sensor arrays for chemical detection include the demonstration of reproducible fabrication methods and functionalization schemes that provide high chemical diversity to the resulting sensors. Here, we outline a scalable approach to fabricating arrays of vapor sensors consisting of NT field effect transistors functionalized with single-stranded DNA (DNA-NT). DNA-NT sensors were highly reproducible, with responses that could be described through equilibrium thermodynamics. Target analytes were detected even in large backgrounds of volatile interferents. DNA-NT sensors were able to discriminate between highly similar molecules, including structural isomers and enantiomers. The sensors were also able to detect subtle variations in complex vapors, including mixtures of structural isomers and mixtures of many volatile organic compounds characteristic of humans.

Carbon nanotube (NT)-based sensors have been demonstrated for environmental, defense and medical sensing applications[1] using various device configurations, including mechanical resonators,[2,3] optical[4] and capacitive sensors,[5,6] and field effect transistors (FETs).[7,8] Chemical modification of the NT surface is a powerful method to influence the interaction strength between the NT and analyte molecules and thereby improve the device sensitivity and specificity. Surface functionalization has been accomplished in many ways, including polymer coatings,[9,10] atomic doping,[11] decoration with metals[12,13] or metal oxides,[14] and coating with single-stranded DNA (DNA).[8,15] The promise of DNA as a functionalizing agent is based on its complex but completely controlled chemistry, which provides affinity for a wide variety of analytes and enables control of sensor responses through choice of the DNA sequence. DNA is available commercially and is sufficiently cheap for use in scalable device fabrication processes. In previous work,[8,15] DNA-NT transistors based on individual, CVD-grown nanotubes were used to detect single analytes at concentrations as low as a few ppb and to distinguish between highly similar compounds, including structural isomers and enantiomers. Responses were fast (seconds), fully reversible, and depended on the identities of the analyte and DNA sequence. Scaling to large arrays was problematic because of the randomness of NT growth, which produces both semiconducting and metallic NTs in a 2:1 ratio, where only the former yield DNA-NT sensors with detectable responses to chemical vapors.[15]

The development of reproducible and scalable fabrication methods for large arrays of NT-based chemical sensors would mark a major step on the path to application of these technologies. Here we report the fabrication of NT FET arrays using commercially available solutions enriched in semiconducting NTs and NT-compatible photolithographic fabrication methods.[16] Arrays of NT FETs had very good device-to-device reproducibility and 90 % yield of useful devices. The arrays were then functionalized to give DNA-NT devices that were tested against compounds characteristically emitted by humans, including a reported volatile marker of skin cancer, as well as sets of molecules with similar chemical structures: *i.e.*, structural isomers and enantiomers. DNA-NT sensors demonstrated highly favorable sensing properties, very similar to those reported for sensors based on single NTs. [8,15] They showed reproducible responses to single analytes that could be fit to predictions from equilibrium thermodynamics. These responses were almost identical when the target was presented in a background with a high concentration of compounds known to block human olfaction. DNA-NT sensors were found to provide differential responses to highly similar compounds, including enantiomers of limonene, and three distinct forms of pinene, a compound with two structural isomers, each with two enantiomeric forms. DNA-NT devices were also tested against vapor mixtures to provide a more realistic assessment of their potential for use in complex environments and medical diagnostics based on volatile biomarkers. The sensors were found to respond to complex mixtures of volatiles characteristically emitted by humans and to be sensitive to slight alterations of the mixture.

## Results and Discussion

DNA-NT vapor sensors were fabricated using a scalable process based on commercial NT solutions, as described in the Methods section. The approach was adapted from earlier reports.[17, 18] Care was taken to develop a method that ensured the production of reproducible arrays of NT transistors where the channel consisted of a relatively sparse NT network (see Fig. 1a). To promote adhesion of NTs to the substrate, a reproducible, uniform monolayer of 3-aminopropyltriethoxysilane (APTES) was deposited using atomic layer deposition (Savannah 200, Cambridge Nanotech), with surface pretreatment by introduction of $H_2O$ vapor to increase the concentration of hydroxyl groups. Optimum values of the concentration of the NT solution and the incubation time were found to be 10 mg/mL and 20 min, respectively. Atomic force microscopy (AFM) of a typical device before DNA deposition (Fig. 1a) showed a sparse network of NTs, typically 1-3 μm long and 0.5-1.5 nm in diameter, that provided multiple conducting pathways connecting the electrodes. The Raman spectrum of deposited NT films (Supplemental Figure 1 of the Supporting Information) showed a high ratio of the intensities of the G and D bands (G/D ~ 50), indicative of a very low defect density. Additionally, the G band was split into two sharp peaks (G- and G+), confirming the high percentage of semiconducting NTs in the sample.[19] As seen in Fig. 1b, current-gate voltage characteristics (I-Vg) of the devices were low-noise with good semiconducting behavior (95% functional device yield with 90% having an on/off ratio exceeding 20; see Fig. 1c). The measured distribution of threshold voltages across devices in a typical array (1.5 V ± 1.8 V) was indicative of low doping and high process reproducibility. A histogram of on-state resistances showed a peak in the range of 1 MΩ; 70% of devices had on-state resistances between 400 kΩ and 4 MΩ (Fig. 1d). These observations were consistent with the expectation that a NT network with multiply connected pathways across the 10 μm channel length has minimal likelihood of a metallic pathway and provides reproducibility by averaging over variations in the NT and substrate properties.

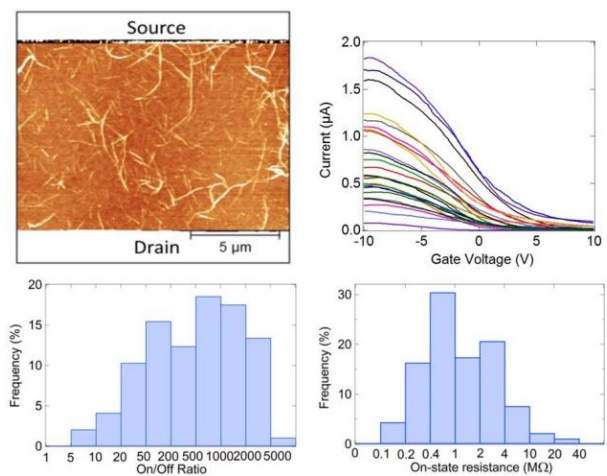

**Figure 1. a)** AFM image of a typical device showing a sparse nanotube network between electrodes. Z-scale is 4 nm. **b)** $I(V_g)$ curves of a representative set of 25 devices, with $V_{DS}$ = 100 mV. **c)** Histogram of on/off ratios shows consistent semiconducting behavior and large on/off ratios. **d)** On-state resistance histogram shows a tight spread, implying good reproducibility across devices.

Electrostatic "chemical gating" has been shown to be a primary mechanism in determining the electrical properties of chemically functionalized NT FETs.[20] It was therefore expected that the magnitude of the change in device current upon exposure to analytes would be proportional to the transconductance of the device, which was typically proportional to the on-state current (Fig. 1b). This justified the use of the normalized change in current, $\Delta I/I_0$, as the sensor response parameter, as used previously for single-NT devices.[15, 21] Other detection mechanisms include changes in carrier scattering and capacitive effects due to adsorbed species. All-atom Molecular Dynamics simulations of DNA-NT indicate that DNA is strongly bound to the NT sidewall by attractive π-π stacking interactions,[22, 23] with a significant number of the bases being desorbed.[24] Our hypothesis is that that this results in a complex, sequence-specific set of binding pockets located within a few nanometers of the NT sidewall. Analyte molecules are solvated by the DNA hydration layer and then bound in the pockets, resulting in the observed DNA-NT signal.

**Table 1: DNA Oligomers Used in the Experiments**

| Name | DNA sequence |
|---|---|
| Seq1 | 5' GAG TCT GTG GAG GAG GTA GTC 3' |
| Seq2 | 5' CTT CTG TCT TGA TGT TTG TCA AAC 3' |
| Seq3 | 5' GCG CAT TGG GTA TCT CGC CCG GCT 3' |
| Seq4 | 5' CCC GTT GGT ATG GGA GTT GAG TGC 3' |

Four different DNA oligomers were used in this work (Table 1). To confirm the formation of a nanoscale DNA layer on the NTs, AFM images were taken of the same region of a NT film before and after DNA functionalization. The data showed a height increase of 0.56 ± 0.2 nm after application of DNA, similar to previous measurements of self-assembled DNA layers on graphene sheets[25] (see Supplemental Figure 2 of the Supporting Information). The substrate height and roughness remained unchanged, indicating DNA deposition was predominantly onto the NTs.

First experiments explored sensor responses to single compounds that are components of human body odors. Dimethylsulfone (DMSO2) is a compound found in human body fluids including skin secretions and volatiles collected above human skin.[26, 27] It has no apparent odor and has been preliminarily identified as a volatile biomarker of basal cell carcinoma.[28] Isovaleric acid is a component of human sweat with an unpleasant odor;[28, 29] it is an unusual compound in that the limit of detection can differ by as much as a factor of 10,000 between individuals due to genetic variation.[30] Fig. 2a shows the responses of five DNA-NT devices based on DNA oligomer Seq2 to DMSO2 (red data) and the average response (black data) at concentrations ranging from 48 – 360 ppb. The device responses were rapid (seconds), reproducible in time and across devices, and they returned to baseline upon flushing with clean air without need for sensor refreshing. Average responses as a function of DMSO2 (isovaleric acid) concentration are plotted in Fig. 2b (Fig. 2c) for several DNA oligomers. In both cases the data are well fit by the prediction of a Langmuir-Hill model of analyte binding dynamics,[31]

$$\frac{\Delta I}{I_0} = A \frac{C^n}{C^n + K_a^n} + Z$$

Here, $C$ is the analyte concentration, $A$ is the magnitude of the response when all binding sites are occupied, $K_a$ is the concentration at which half the maximum response is seen and $n$ is the Hill coefficient describing cooperativity of binding. The best fit values for the offset parameter, $Z$, were very small, typically less than 0.1%. For the analytes tested, the best fit values for $K_a$ were typically 0.5 to a few 10s of ppm, and the cooperativity parameter was close to 1 suggesting independent analyte binding. We drew the conclusion that DNA-NT devices were in thermal equilibrium with analyte vapors tested, consistent with the observation of rapid response and recovery (Fig. 2). Fit parameters for responses of DNA-NT based on the various DNA sequences to DMSO2 and isovaleric acid are summarized in Supplemental Table 1 of the Supporting Information. Additional numerical data quantifying sensor responses to these compounds are presented in Supplemental Table 2.

The sign and magnitude of DNA-NT sensor responses depended on both the DNA sequence used and the analyte. Isovaleric acid (pKa = 4.8) produced a positive signal (current increase), consistent with the expectation that it deprotonated and acquired a negative charge in the DNA hydration layer. DMSO2 was readily detected at the level of 10s of ppb. This species was expected to be uncharged in water, so the signal was ascribed to a dipolar interaction, similar to earlier reports.[20] Along with the results of additional experiments presented below, these data suggested that the DNA/NT sensor class has a high degree of chemical diversity, an important requirement for construction of a functional e-nose system.[32, 33]

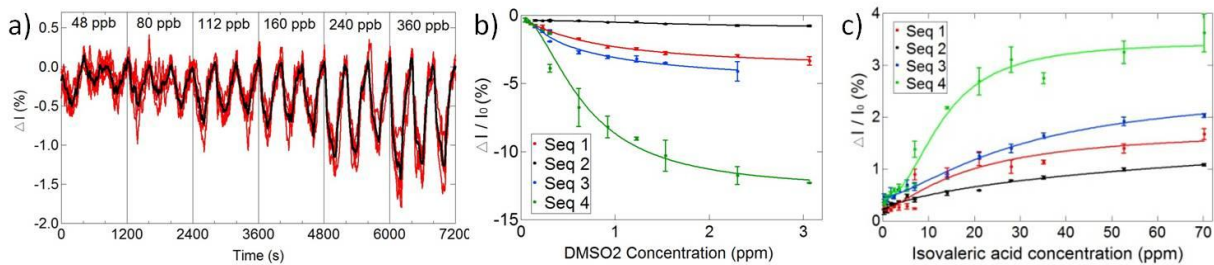

**Figure 2: a)** Responses of five DNA-NT devices based on Seq 4 to repeated pulses of DMSO2, with concentration in the range 48 – 360 ppb (red data). The average response is shown in black. **b)** Average responses of DNA-NT based on four different DNA oligomers to DMSO2 and the corresponding Langmuir-Hill fits. **c)** Similar data and fits for responses of DNA-NT based on four different DNA oligomers to isovaleric acid.

We next tested the ability of DNA-NT sensors to differentiate analytes with very similar molecular structure. Experiments were based on enantiomers of limonene and three isomers of pinene, a compound that has two structural isomers, each of which has a pair of enantiomers. The averaged responses of DNA-NT based on Seq1 to the enantiomers of limonene are shown in Fig. 3a; there is clear discrimination between these highly similar molecules. The standard Langmuir-Hill fit worked well for the data for D(+) limonene but poorly for L(-) limonene. Responses for L(-) limonene showed anomalous behavior with a positive response at low concentrations that crossed over to a negative response for concentrations exceeding ca. 40 ppm. This behavior is consistent with the presence of two distinct types of binding sites for limonene, one that leads to a positive response for both enantiomers and one that leads to positive and negative responses for the D(+) and L(-) enantiomers, respectively. Given that previous experiments from our lab, which employed DNA-NT based on Seq1 and single NTs grown by CVD, also showed positive responses for the D(+) enantiomer and negative responses for the L(-) form,[8] we suggest that the binding site that distinguished between enantiomers was associated with the DNA, while the other site corresponded to binding of the analyte to junctions between NTs in the network.

Isomers of pinene were also clearly distinguished by DNA-NT sensors (Fig. 3b); both the double bond location and the handedness of the pinene molecule affected the sensor responses, with DNA-NT based on Seq3 showing greater differentiation than those based on Seq1. As a further test of the differentiation power, DNA-NT devices based on Seq1 and Seq3 were tested against mixtures of the α(-) and β(-) structural isomers of pinene (Fig. 3c). The averaged responses of 5 devices based on Seq3 provided the ability to resolve composition changes of approximately 5-10% in a vapor where the total pinene concentration was held fixed at 130 ppm. Devices based on Seq1 showed less discrimination power for the mixtures of α(-) and β(-) pinene than those based on Seq 3, consistent with their respective responses to the neat analytes (Fig. 3b). Additional data quantifying the response of DNA/NT based on the four oligomers is found in Supplemental Table 2 of the Supporting Information. From these measurements it was concluded that the conformation of DNA bound to the NT sidewall was sufficiently complex to enable differential binding between these sets of highly similar molecules.

Stereo-specific sensing has been reported for other sensor modalities,[34-36] including DNA-NT sensors based on single CVD-grown NTs.[8] The chiral nature of DNA admits the possibility of stereospecific interactions, which is of particular relevance in understanding the activity of DNA-binding drugs.[37] All-atom Molecular Dynamics simulations provide an approach to understanding the structure of DNA-NT hybrids[22-24] and should prove useful in unraveling the molecular mechanisms of enantiomer discrimination in this system.

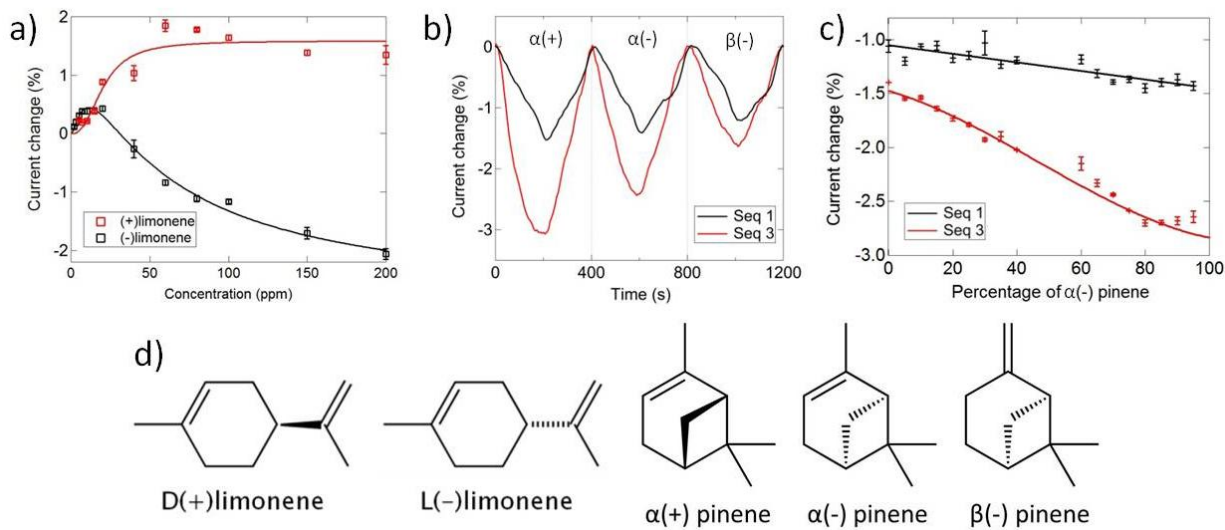

Figure 3: a) DNA-NT devices based on Seq1 clearly distinguish limonene enantiomers. The responses to D(+) limonene (red data) are well fit by a simple Langmuir-Hill equation, while responses to L(-) limonene (black data) require a two-component fit, suggesting the existence of two distinct binding sites. b) Responses of DNA-NT based on Seq 1 (black data) and Seq 3 (red data) to pulses of α(+),α(-) and β(-) pinene at a concentration of 130 ppm. The responses depend on both the location of the double bond and the handedness of the molecule. c) The responses of DNA-NT based on Seq1 and Seq3 decrease as the analyte is adjusted from pure α(-) pinene to an α(-)/β(-) mixture to pure β(-) pinene. d) Chemical structures of the limonene and pinene isomers used in the experiments.

DNA-NT sensor responses to isovaleric acid and its more pleasant-smelling ethyl ester (ethyl isovalerate) were examined to determine if the ester blocked the response of the DNA-NT to the acid. This experiment was based on previous studies, using *in vivo* olfaction, which demonstrated that ethyl esters of an organic acid responsible, in part, for human axillary odor[38] (*e.g.*, E-3-methyl-2-hexenoic acid), reduced the perception of this malodorous compound.[39] Strikingly, responses of DNA-NT sensors based on Seq3 to the target isovaleric acid at concentrations of 7 – 53 ppm in clean air were almost identical to the responses in an interfering background of ethyl isovalerate at 3300 ppm (Fig. 4a).

In a second test, introduction of 460 ppm cis-3-hexen-1-ol left the responses of DNA-NT based on Seq3 to DMSO2 at 0.76 – 6.1 ppm essentially unchanged (Fig. 4b). This is significant because cis-3-hexen-1-ol is known to mask odors from human olfaction and is used in deodorant products by the fragrance industry.[40, 41] The lack of effective blocking of the response in these two cases presumably derives from the fact that the nature of DNA as a "receptor" for volatile compounds differs significantly from that of human olfactory receptor proteins (ORs). Thus, compounds that act to diminish the binding of specific volatiles to human ORs may have little or no effect on responses of DNA-NT devices.

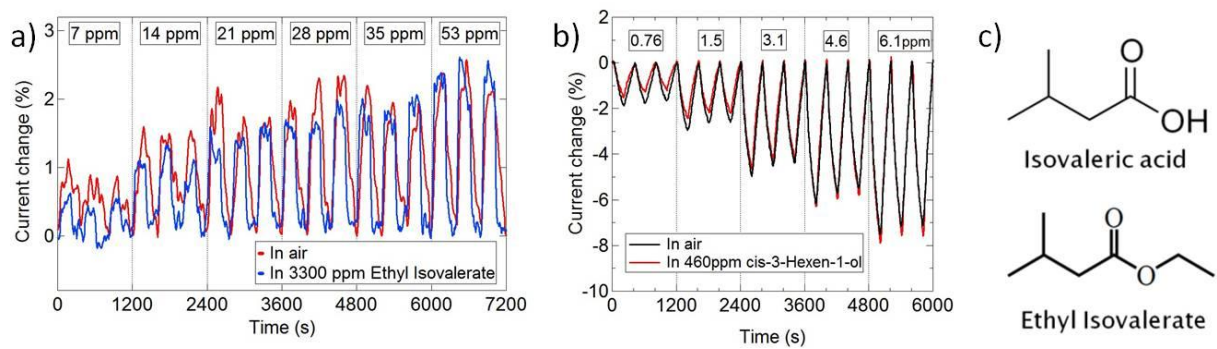

**Figure 4:** a) DNA-NT sensors based on Seq3 show near identical responses to 7-53 ppm isovaleric acid in clean air (red data) and in a background of 3300 ppm ethyl isovalerate (blue data). b) Average responses from four DNA-NT sensors based on Seq3 to 0.76 – 6.1 ppm DMSO2 in clean air (black data) and in a background of 460 ppm cis-3-hexen-1-ol (red data). c) Chemical structures of target analyte isovaleric acid and blocking compound ethyl isovalerate.

An essential characteristic of biological olfactory systems is the ability to differentiate between very similar *complex mixtures* of volatile compounds. Experiments to test the ability of DNA-NT devices in this domain were based on a mixture of 17 organic compounds, many of which are volatile and found in non-axillary skin sweat.[25] These were dissolved in physiological saline (see Table 2). DNA-NT devices were exposed to the headspace vapor of the original "parent" mixture and also to that of "spiked" mixtures where one component was increased in concentration by a factor of 2-10. "Spiked" mixtures were based on compounds that were prevalent in the mixture (acetic acid) and those found in trace amounts (stearic acid and nonanal), with widely varying vapor pressures.

**Table 2: Concentration and Vapor Pressure of Components of the Parent Complex Mixture Used in the Experiments**

| Compound | Concentration, mg/mL | Vapor Pressure, Torr (@ 20 C unless stated) |
|---|---|---|
| Acetic Acid | 0.67 | 3.0 |
| Lactic Acid | 0.66 | 0.08 |
| Glycerol | 0.17 | 1 @ 125 C |
| Stearic Acid | 0.03 | 1 @ 174 C |
| Acetoin | 0.05 | 2.69 |
| Propanoic Acid | 0.09 | 2.9 |
| Isobutyric Acid | 0.01 | 1.5 |
| Butyric Acid | 0.45 | 0.43 |
| Isovaleric Acid | 0.01 | 0.38 |
| 2-Methylbutyric Acid | 0.01 | 0.5 |
| Isocaproic Acid | 0.01 | N/A |
| 4-Methyl-phenol | 0.05 | 1 |
| Phenol | 0.01 | 0.36 |
| Dimethylsulfone | 0.05 | N/A |
| Nonanal | 0.01 | 0.26 |
| Indole | 0.03 | 0.03 |
| Squalene | 0.20 | 2 @ 240 C |

The concentrations of various components in the headspace of the mixtures were not measured. However, estimates were made using Raoult's law, which assumes that the concentration of a mixture component is the product of the vapor pressure of the component and its molar fraction in the solution. For the parent mixture this yielded 793 ppb for acetic acid and 0.43 ppb for nonanal. Stearic acid is a solid at room temperature with a very low vapor pressure (see Table 2). Although no precise estimate could be formulated, the expected concentration in the headspace of the parent mixture would surely be well below 1 ppb. For a spiked mixture, these concentrations were multiplied by the appropriate spiking factor (*i.e.*, 2x, 5x, or 10x).

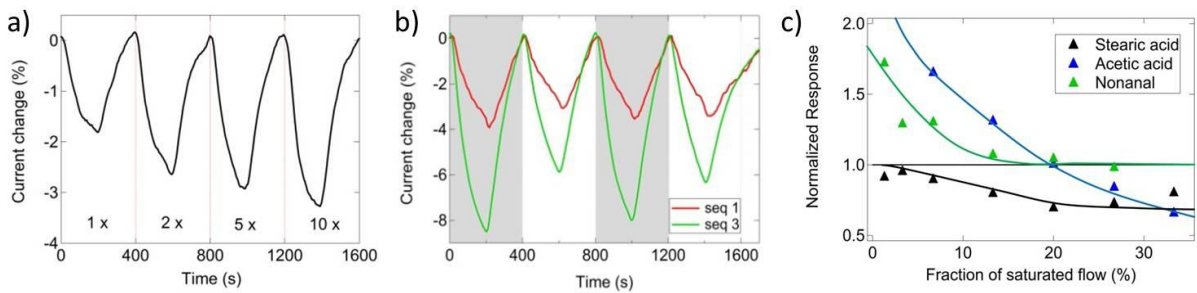

**Figure 5: a) DNA-NT device based on Seq3 provides clear differential responses between the "parent" mixture and mixtures "spiked" with nonanal by factors of two, five and ten (2x, 5x, 10x, respectively). b) Responses of DNA-NT based on Seq1 (red data) and Seq3 (green data) to 33% saturated vapor of the "parent" mixture (gray background) and a "spiked" mixture with 10x increased concentration of stearic acid (white background). Devices based on Seq3 show a strong differential signal to the two mixtures while the differential signal for DNA-NT based on Seq1 is weak. c) Responses of DNA-NT based on Seq4 to diluted streams of headspace vapor of "spiked" mixtures, normalized to the response to the parent mixture. In each case, the named component is spiked by a factor 10x compared to its concentration in the parent solution. Markers are experimental data, and the solid lines are guides to the eye.**

The spiked mixtures were exposed to DNA-NT devices concurrently with the standard mixture and the sensor responses ($\Delta I/I_0$) were recorded. DNA-NT devices were typically, but not always, found to provide strong differential responses between the parent mixture and spiked mixtures. Responses of DNA-NT based on Seq3 were very sensitive to the concentration of nonanal in the mixture (Fig. 5a). We note that the estimated concentration of nonanal in the vapor ranged from 0.43 – 4.3 ppb. Differential response of DNA-NT devices to the "parent" and "spiked" mixtures depended on the identity of the DNA oligomer. For example, responses of DNA-NT based on Seq1 to the "parent" mixture and a mixture "spiked" by 10x with stearic acid were nearly identical, while responses of DNA-NT based on Seq3 showed clear differentiation between these two mixtures (Fig. 5b, red and green data, respectively). The concentration of stearic acid in the vapor is not precisely known but is almost certainly at the level of a few ppb or lower. The implication is that stearic acid does not significantly bind to Seq1 in the presence of all the other VOCs but that it does bind strongly to sequence 3. Furthermore, a very rich data set was obtained by considering DNA-NT differential responses to the headspace vapor of the "spiked" mixture at various dilutions with clean air. As seen in Fig. 5c, the dependence of the differential responses of DNA-NT based on Seq4 with dilution depends on the identity of the "spiked" component. Consequently, real world mixtures could potentially be identified by comparison to a standard mixture using DNA-NT sensors. By measuring response deviation as a function of dilution, it could be possible to identify exactly which compound in the mixture has been altered and by how much its concentration has changed.

## Conclusions

In summary we have demonstrated a facile, potentially scalable method for fabricating DNA-NT vapor sensors that could enable their use in sensor arrays suitable for incorporation into an electronic nose system. We tested device responses against individual VOC analytes characteristic of humans and against complex mixtures that more closely resemble "real-world" samples. DNA-NT sensors showed excellent reproducibility, they responded within seconds to parts per billion concentrations, and their responses were in good agreement with predictions of equilibrium thermodynamics. Devices were able to differentiate between analytes with very similar molecular structure (*i.e.*, enantiomers and structural isomers), they were able to detect target analytes in a large background of an interfering VOC, and they could discriminate subtle changes in complex VOC mixtures. The use of DNA as the functionalizing agent holds the possibility that arrays of hundreds or thousands of individual sensors could be fabricated on a single chip and functionalized with a large number of different DNA oligomers. As each sequence has its own set of characteristic responses to a large number of various analytes, this approach should allow analytes to be detected and distinguished at relevant concentrations across many applications, including deducing chemical composition in an unknown environment or disease diagnosis from VOCs.

## Materials and Methods

**Device Fabrication and Functionalization:** Electrical contacts for FETs with channels 10 μm long and 25 μm wide were patterned by photolithography and metallized with Cr/Au *via* thermal evaporation. After $O_2$ plasma cleaning to remove residual photoresist, a 3-aminopropyltriethoxysilane (APTES) monolayer was deposited using atomic layer deposition (Savannah 200, Cambridge Nanotech), with surface pretreatment by introduction of $H_2O$ vapor to increase the concentration of hydroxyl groups. Semiconducting NTs were deposited from solution (NanoIntegris, Isonanotubes-S 98%) by pipetting onto the surface of the chip and incubation in a humid atmosphere for 20 minutes. The NT-FET arrays were cleaned by immersion in isopropanol followed by immersion in a deionized (DI) water bath, followed annealing at 200°C for 1 hour to improve the electrical contacts. DNA solutions with a concentration of 100 μM were prepared by adding deionized (DI) water to as-received DNA (Invitrogen). Devices were functionalized by incubation in droplets of DNA solution in a humid atmosphere to suppress droplet evaporation. After 30 minutes, the DNA solution was blown off the wafer with compressed nitrogen, taking care not to cross-contaminate devices with other DNA sequences.

**Measurement of Device Responses to Analyte Vapors:** Analyte vapors were delivered into a home-built 6 cm x 2.5 cm x 1 cm chamber using a computer-controlled system of mass flow controllers. Clean air (Praxair UN1002) was flowed through a bubbler containing analyte liquid to create a stream of saturated vapor (flow rate of 1 – 500 sccm) and through a bubbler containing water to create a stream of saturated water vapor (500 sccm). These two streams were mixed with a stream of clean air whose flow rate was controlled such that all measurements were taken with a constant total flow rate of 1500 sccm, at 33% relative humidity. Devices were electrically contacted using feed-through connections into the chamber. A bias voltage (typically 100 mV) was applied to the devices and currents of up to 10 devices were read out sequentially during the same run using a switching matrix (Keithley 7001) and picoammeter (Keithley 6485). Between 5 and 20 devices were measured for each DNA-analyte combination and the sensing response was defined as the normalized change in the conductance, averaged across multiple devices and measurements.

**Preparation of Chemical Mixtures:** A 5000 ppm solution of dimethylsulfone (DMSO2; Cambridge Isotope Laboratories, Inc.) in dipropylene glycol was prepared by heating and sonicating. The headspace concentration was measured by gas chromatography-mass spectrometry, as previously described,[8] and found to be 23 ppm. Other analytes used were purchased from Sigma Aldrich and used as received. Mixtures of analytes were made by diluting in deionized water and shaking and sonicating as required to dissolve.

# Acknowledgements

We acknowledge useful discussions with Dr. E. Dattoli. This research was supported by the Department of Defense US Air Force Research Laboratory and UES through Contract Nos. FA8650-09-D-5037 as well as support from the Army Research Office through grant #W911NF-11-1-0087. M.L. acknowledges the support of a SMART Fellowship. N.J.K. and J.S.Y. acknowledge the support of the Nano/Bio Interface Center through the National Science Foundation NSEC DMR08-32802. Use of the facilities of the Nano/Bio Interface Center is also acknowledged.

*Supporting Information Available*: Raman spectroscopy data for carbon nanotube films, indicating their low defect density and high content of semiconducting nanotubes. AFM data and associated histogram used to determine the distribution of diameters of the carbon nanotubes. Tables showing 1) details of the Langmuir-Hill fits for four different DNA sequences and two vapor analytes (DMSO2 and isovaleric acid); 2) response data for analyte-DNA oligomer combinations tested; and 3) response data for DNA-NT based on various DNA oligomers tested against the "parent" and "spiked" mixtures mentioned in the main text. This information is available free of charge *via* the Internet at http://pubs.acs.org.